\documentclass[12pt] {article}
\begin{document}
\mbox{}
\begin{center}
{\Large \bf
Theory of Doping and Defects in III-V Nitrides}
\\[0.5cm]
Chris G. VAN DE WALLE,$^a$ Catherine STAMPFL,$^a$ and J\"{o}rg
NEUGEBAUER$^b$\\
$^a$Xerox Palo Alto Research Center, Palo Alto, California 94304, USA\\
Tel: +1-650-812-4163; Fax +1-650-812-4140; e-mail:
vandewalle@parc.xerox.com\\
$^b$Fritz-Haber-Institut, Abt.\ Theorie, Faradayweg 4--6, D-14195
Berlin, Germany
\\[0.8cm]
\end{center}

Doping problems in GaN and in AlGaN alloys are addressed on the basis
of state-of-the-art first-principles calculations.  For $n$-type doping
we find that nitrogen vacancies are too high in energy to be
incorporated during growth, but silicon and oxygen readily form
donors.  The properties of oxygen, including $DX$-center formation,
support it as the main cause of unintentional $n$-type conductivity.
For $p$-type doping we find that the solubility of Mg is the main
factor limiting the hole concentration in GaN.  We discuss the
beneficial effects of hydrogen during acceptor doping.  Compensation of
acceptors by nitrogen vacancies may occur, becoming increasingly severe
as $x$ increases in Al$_{x}$Ga$_{1-x}$N alloys.
\\[0.5cm]
{\bf 1. Introduction}
\\[-0.1cm]

Tremendous progress has recently been made in the growth and
fabrication of GaN-based electronic and optoelectronic devices
\cite{poncebour}.  A number of problems still exist, however, which may
hamper further progress.  Control of doping levels is essential; for
$n$-type doping, the outstanding problems include (i) suppression of
background $n$-type conductivity; (ii) compensation at high doping
levels; and (iii) $n$-type doping of AlGaN alloys.   For $p$-type
doping, the main issue concerns increasing the achievable hole
concentrations; this requires an understanding of (i) compensation by
native defects; (ii) potential metastabilities (as evidenced in
persistent photoconductivity); (iii) the role of hydrogen; and (iv) the
reasons for increased doping difficulties in AlGaN alloys.  In this
paper we discuss how a theoretical approach for native defects and
dopant impurities, combined with state-of-the-art first-principles
calculations, can be used to understand the various factors that govern
doping.

After a brief description of the theoretical methods in Section 2, we
will discuss our results for $n$-type doping in Section 3.  We will
show that nitrogen vacancies play no role in $n$-type conductivity, and
that unintentional $n$-type doping is usually due to oxygen.  We
discuss the behavior of oxygen in AlGaN alloys, where a $DX$ transition
is predicted to occur.  Silicon donors do not exhibit this transition,
and it is also absent in the zinc-blende phase.  We will also discuss
gallium vacancies, which act as compensating centers in $n$-type GaN,
and which are the most likely source of the ``yellow luminescence.''

Section 4 will address $p$-type doping, which is now routinely
performed in GaN with Mg acceptors and using a post-growth activation
step in material grown by MOCVD (metal-organic chemical vapor
deposition).  Hydrogen plays a crucial role in this process.  However,
$p$-type doping levels are still lower than desirable for
low-resistance cladding layers and ohmic contacts.  We will show that
Mg solubility is the determining factor limiting the hole concentration
in GaN; incorporation of Mg on interstitial sites or on substitutional
nitrogen sites is found to be unfavorable.   We will also discuss the
prospects of other acceptor impurities for achieving higher doping
levels.  Some degree of compensation by nitrogen vacancies occurs, and
we will discuss the effect of increasing $x$ in Al$_{x}$Ga$_{1-x}$N
alloys on the doping efficiency.  Section 5 concludes the paper.
\\[0.5cm]
{\bf 2. Methods}
\\[-0.1cm]

The equilibrium concentration of impurities or native defects can be
expressed as
\begin{equation}\label{eq:def_conc}
c = N_{\rm sites}  \exp^{-E^f/k_B\,T}
\end{equation}
Where $E^f$ is the {\it formation energy}, $N_{\rm sites}$ is the
number of sites the defect or impurity can be incorporated on, $k_B$ is
the Boltzmann constant, and $T$ the temperature.  Equation
(\ref{eq:def_conc}) shows that defects with a {\it high} formation
energy will occur in {\it low} concentrations.

The formation energy is not a constant but depends on the growth
conditions. For example, the formation energy of an oxygen donor is
determined by the relative abundance of O, Ga, and N atoms, as
expressed by the chemical potentials $\mu_{\rm O}$, $\mu_{\rm Ga}$ and
$\mu_{\rm N}$, respectively.  If the O donor is charged (as is expected
when it has donated its electron), the formation energy depends further
on the Fermi level ($E_F$), which acts as a reservoir for electrons.
Forming a substitutional O donor requires the removal of one N atom and
the addition of one O atom; the formation energy is therefore:
\begin{equation}\label{eq:form}
E^f({\rm \mbox{GaN:O}}_{\rm N}^q) = E_{\rm tot}({\rm \mbox{GaN:O}}_
  {\rm N}^q)
    - \mu_{\rm O} + \mu_{\rm N} + q E_F
\end{equation}
where $E_{\rm tot}({\rm \mbox{GaN:O}}_{\rm N}^q)$ is the total energy
derived from a calculation for substitutional O, and $q$ is the charge
state of the O donor.  Similar expressions
apply to other impurities and to the various native defects.  We refer
to Refs.~\cite{prb1_94} and \cite{festk96} for a more
complete discussion of formation energies and their dependence on
chemical potentials.

The quantity $E_{\rm tot}$ in Eq.~(\ref{eq:form}) is obtained from
state-of-the-art first-principles calculations that do not require any
adjustable parameters or any input from experiment.  The computations
are founded on density-functional theory, using a supercell geometry
and soft {\it ab initio} pseudopotentials.  The effect of $d$ electrons
in GaN is taken into account either through the so-called non-linear
core correction or by explicit inclusion of the $d$ electrons as
valence electrons; the latter proved to be necessary for obtaining
accurate results in certain cases \cite{mrs94}.  References and further
details of the computational approach can be found in
Refs.~\cite{prb1_94}, \cite{stumpf94}, and \cite{mrs95_tb}.

The Fermi level $E_F$ is not an independent parameter, but is
determined by the condition of charge neutrality.  In principle we can
write equations such as (\ref{eq:form}) for every native defect and
impurity in the material, and then solve the complete problem
(including free-carrier concentrations in valence and conduction bands)
self-consistently, imposing charge neutrality.  However, it is
instructive to plot formation energies as a function of $E_F$ in order
to examine the behavior of defects and impurities when the doping level
changes.  As for the chemical potentials, these are variables which
depend on the details of the growth conditions.  For clarity of
presentation, we set these chemical potentials to fixed values in the
figures shown below; however, a general case can always be addressed by
referring back to Eq.~(\ref{eq:form}).  The fixed values we have chosen
correspond to Ga-rich conditions [$\mu_{\rm Ga} = \mu_{\rm Ga(bulk)}$],
and to maximum incorporation of the various impurities, with
solubilities determined by equilibrium with Ga$_2$O$_3$, Si$_3$N$_4$,
and Mg$_3$N$_2$.
\\[0.5cm]
{\bf 3. $n$-type doping}
\\[-0.1cm]

For a long time the nitrogen vacancy was thought to be the source of
$n$-type conductivity in GaN.  Our first-principles investigations,
however, indicate that nitrogen vacancies are high-energy defects in
$n$-type GaN \cite{prb1_94}.  Nitrogen vacancies ($V_{\rm N}$) do
behave as donors; if they are purposely created, for instance during
ion implantation, they will increase the electron concentration.
However, their high formation energy makes it very unlikely that they
would form during growth of undoped or $n$-type GaN, and hence they
cannot be responsible for $n$-type conductivity.
Instead, we have proposed that unintentional impurities such as oxygen
and silicon are the actual cause of the observed unintentional $n$-type
doping \cite{icds_94}.  These impurities are calculated to be shallow
donors with high solubilities.  Figure 1 summarizes our results for
native defects and impurities relevant for $n$-type doping.  It is
clear that O and Si have much lower formation energies than $V_{\rm
N}$; these impurities can be readily incorporated in $n$-type GaN.
Both O and Si form shallow donors in GaN.  The slope of the lines in
Fig.~1 indicates the charge state of the defect or impurity [see
Eq.~(\ref{eq:form})]:  Si$_{\rm Ga}$, O$_{\rm N}$, and $V_{\rm N}$ all
appear with slope +1, indicating single donors.

\begin{picture}(20,90)(20,100)
\put(-150,-330){\includegraphics{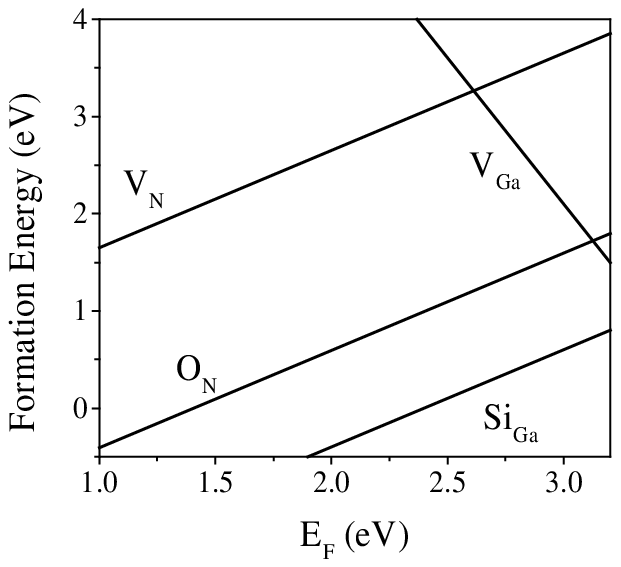}}
\end{picture}
\hspace{10.0cm}\parbox{2.5in}{
\vspace{1.4cm}
{\it Figure 1: Formation energy {\it vs.} Fermi energy for native
defects (nitrogen and gallium vacancies) and donors (oxygen and
silicon) in GaN.  The zero of Fermi energy is located at the top of the
valence band.}}
\vspace{1.0cm}

A few reports had proposed oxygen as a potential source of $n$-type
conductivity in GaN \cite {seifert83,chung92}.  Still, the prevailing
conventional wisdom, attributing the $n$-type behavior to nitrogen
vacancies, proved hard to overcome.  Recent experiments have now
confirmed that unintentionally doped $n$-type GaN samples contain
Si or O concentrations high enough to explain the electron
concentrations \cite{gotz96_1,gotzmrs96}.  High levels of $n$-type
conductivity have always been found in GaN bulk crystals grown at high
temperature and high pressure \cite{perlin95}.  Recent high-pressure
studies have established that the characteristics of these samples are
very similar to epitaxial films which are intentionally doped with
oxygen \cite{wetzel96,perlin96}.  The $n$-type conductivity of bulk GaN
can therefore be attributed to unintentional oxygen incorporation.  The
high-pressure experiments have also shown that freezeout of carriers
occurs at pressures exceeding 20 GPa
\cite{perlin95,perlin96,wetzel97}.  Originally this observation was
interpreted as consistent with the presence of nitrogen vacancies,
since the $V_{\rm N}$ donor gives rise to a resonance in the conduction
band, which emerges into the band gap under pressure.  However, the
observations are also entirely consistent with a ``$DX$-like'' behavior
of the oxygen donor.

The prototype $DX$ center is Si in GaAs, which undergoes a
transition from a shallow to a deep center when hydrostatic pressure is
applied, or in AlGaAs alloys \cite{mooney}.  The transition is
accompanied by a strong relaxation of the impurity off the
substitutional site \cite{chadichang88}.  We have carried out extensive
calculations for oxygen in GaN under pressure, as well as in AlGaN
alloys \cite{vdwun}.  Under compression the oxygen impurity assumes an
off-center configuration: a large outward relaxation introduces a deep
level in the band gap.  This behavior explains the carrier freezeout in
GaN under pressure.  Silicon donors do not exhibit this transition,
consistent with experiment \cite{wetzel97}.  Alloying with AlN
increases the band gap similar to the application of hydrostatic
pressure; one therefore expects that the behavior of the impurities in
AlGaN would be similar to that in GaN under pressure.  Explicit
calculations for oxygen in AlN indeed produce $DX$ behavior
\cite{vdwun}.  These results are consistent with the observed decrease
in $n$-type conductivity of unintentionally doped Al$_x$Ga$_{1-x}$N as
$x>0.4$ \cite{leegershenzon91}.

Interestingly, we find that the $DX$ transition does {\it not} occur in
{\it zinc-blende} (ZB) AlGaN.  This difference is surprising, since the
local environment of the impurity is very similar (the two phases only
differ at the positions of third-nearest neighbors and beyond, and no
qualitative differences had been observed for any defects or impurities
so far \cite{mrs94,prl95_hy}.  We have explained the difference in $DX$
behavior by analyzing the interaction between the oxygen impurity and
third-nearest neighbors, showing that a Coulombic attraction provides a
driving force for the large lattice relaxation \cite{vdwun}.

Finally, we note in Fig.~1  that gallium vacancies
($V_{\rm Ga}^{3-}$) have relatively low formation energies in highly
doped $n$-type material ($E_F$ high in the gap); they could therefore
act as compensating centers.  Yi and Wessels \cite{wessels96} have
found evidence of compensation by a triply charged defect in Se-doped
GaN.  We  have also proposed that gallium vacancies are responsible for
the ``yellow luminescence'' (YL) in GaN, a broad luminescence band
centered around 2.2~eV \cite{apl95_yl}.  The origins of the
YL have been extensively debated; as discussed in
Refs.~\cite{apl95_yl} and \cite{vdwmrsfall96}, the calculated
properties of the gallium vacancy are in good agreement with
experimental results.
\\[0.5cm]
{\bf 4. $p$-type doping}
\\[-0.1cm]

$p$-type doping levels in GaN and AlGaN alloys are still lower than
desirable for low-resistance cladding layers and ohmic contacts.
Achieving higher hole concentrations with Mg as the dopant has proved
difficult.  Our investigations \cite{mrs_95} have revealed that the
determining factor is the solubility of Mg in GaN, which is limited by
competition between incorporation of Mg acceptors and formation of
Mg$_3$N$_2$.  It would be interesting to investigate experimentally
whether traces of Mg$_3$N$_2$ can be found in highly Mg-doped GaN.  Mg
prefers the substitutional Ga site, and incorporation of Mg on
substitutional N sites (Mg$_{\rm N}$) or on interstitial sites (Mg$_i$)
was found to be unfavorable.  These features are illustrated in
Fig.~2.

\begin{picture}(20,90)(20,100)
\put(-150,-360){\includegraphics{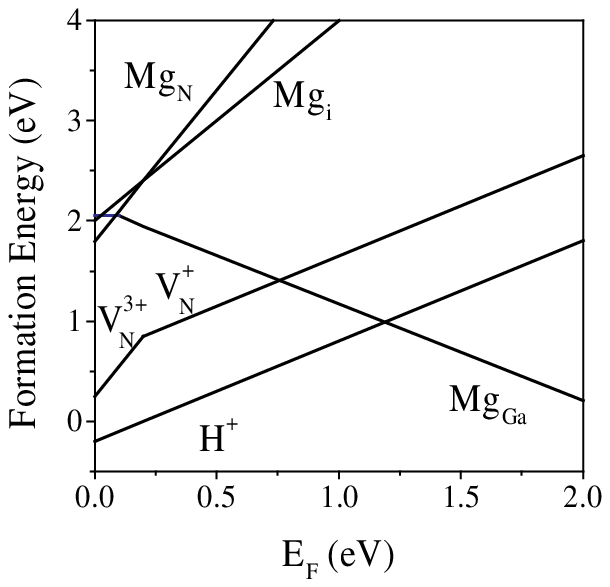}}
\end{picture}
\hspace{10.0cm}\parbox{2.5in}{
\vspace{1.4cm}
\it Figure 2: Formation energy as a function of Fermi level for Mg in
different configurations (Ga-substitutional, N-substitutional, and
interstitial configuration). Also included are the native defects and
interstitial H.}
\vspace{1.0cm}

Hydrogen has strong effects on the properties of $p$-type GaN.  Many
growth techniques, such as MOCVD or hydride vapor phase epitaxy (HVPE)
introduce large concentrations of hydrogen in the growing material.
The behavior of hydrogen in GaN was analyzed in detail in
Refs.~\cite{prl95_hy} and \cite{apl95_hy}.  We found that hydrogen
incorporates much more readily in $p$-type than in $n$-type GaN.  In
$p$-type GaN H behaves as a donor (H$^+$), compensating acceptors.
Hydrogen can bind to the Mg acceptors with a binding energy of 0.7 eV.
The structure of the resulting complex is unusual in that H does not
sit next to the Mg, but binds to a N atom which is a neighbor of the
acceptor.  As a direct consequence the vibrational frequency of the
complex is {\it not} representative of a Ga-H bond, but rather of a N-H
bond.  The calculated vibrational frequency (in the harmonic
approximation) is 3360 cm$^{-1}$.  Anharmonic effects may lower this
frequency by as much as 170 cm$^{-1}$ \cite{johnson93}.  G\"{o}tz {\it
et al.} \cite{gotz96} have reported a value of 3125 cm$^{-1}$ for this
local vibrational mode.

The nitrogen vacancy, which had a high formation energy in $n$-type GaN
(see Fig.~1), has a significantly lower formation energy in $p$-type
material, and could act as a compensating center.  Figure 2 shows that
$V_{\rm N}$ can occur in a 3+ as well as a + charge state; the +/3+
transition is characterized by a large lattice relaxation
\cite{festk96}.  Compensation by $V_{\rm N}$ may therefore be
responsible for the observed persistent photoconductivity effects
\cite{johnsonli96}.  The metastability is associated with the different
position of the $A_1$ state near the valence band in the +1 and +3
charge states; this state is occupied with two electrons the +1 charge
state, and empty for the 3+ charge state.  The nitrogen vacancy also
may give rise to the blue lines (around 2.9 eV) commonly observed by
photoluminescence in Mg-doped GaN.

We have carried out similar calculations for the formation energy of
native defects and impurities in AlN.  The nitrogen vacancy has a
strikingly lower formation energy in AlN than in GaN; it will therefore
introduce more severe compensation.  We propose that compensation by
nitrogen vacancies is the likely cause of the decreased doping
efficiency of Mg when the Al content is raised in Al$_{x}$Ga$_{1-x}$N
alloys.

Figure 2 also shows that hydrogen, when present, has a formation
energy much lower than that of the nitrogen vacancy.  In growth
situations where hydrogen is present (such as MOCVD or HVPE) Mg-doped
material will preferentially be compensated by hydrogen, and
compensation by nitrogen vacancies will be suppressed.  The presence of
hydrogen is therefore beneficial -- at the expense, of course, of
obtaining material that is heavily compensated by hydrogen!
Fortunately, the hydrogen can be removed from the active region by
treatments such as low-energy electron-beam irradiation \cite{amano89}
or thermal annealing \cite{nakamura92}.  A more complete discussion of
the role of hydrogen in GaN is given in Refs.~\cite{prl95_hy} and
\cite{apl95_hy}.

For Mg, we thus conclude that achievable doping levels are mainly
limited by the solubility of Mg in GaN.  We have investigated other
candidate acceptors, and evaluated them in terms of solubility, shallow
{\it vs.}~deep character, and potential compensation due to
incorporation on other sites \cite{icps96}.  None of the candidates
exhibited characteristics exceeding those of Mg.  In particular, we
perceive no noticeable advantage in the use of Be, which has been
suggested as a superior dopant.

Finally, we note the importance of avoiding oxygen contamination during
growth of $p$-type GaN.  The oxygen formation energy shown in Fig.~1
clearly extrapolates to very low values in $p$-type GaN.  Any oxygen
present in the growth system will therefore be readily incorporated
during $p$-type growth.  In addition, complex formation between oxygen
and magnesium can make oxygen incorporation even more favorable: we
find that the Mg-O complex has a binding energy of 0.6 eV in both GaN and
AlN.
\\[0.5cm]
{\bf 5. Conclusions}
\\[-0.1cm]

We have presented a comprehensive overview of our current understanding
of point defects and dopant impurities in nitride semiconductors, based
on first-principles calculations.  Our main conclusions for $n$-type
GaN are that (i) nitrogen vacancies are {\it not} responsible for
unintentional $n$-type conductivity; (ii) Si and O donors can be
incorporated in large concentrations, likely causing unintentional
$n$-type doping; (iii) oxygen (but not silicon) behaves as a $DX$
center in GaN under pressure and in AlGaN alloys; (iv) the $DX$
transition does not occur in zinc-blende material; and (v) gallium
vacancies are the likely source of the yellow luminescence.  For
$p$-type GaN we found that (i) the hole concentration obtained with Mg
doping is limited due to Mg solubility; (ii) incorporation of Mg on
interstitial sites or antisites is not a problem; (iii) hydrogen has a
beneficial effect on $p$-type doping because it suppresses compensation
and enhances acceptor incorporation; (iv) compensation by nitrogen
vacancies may occur, becoming worse as $x$ increases in
Al$_{x}$Ga$_{1-x}$N; and (v) other acceptor impurities do not exhibit
characteristics superior to Mg.
\\[0.5cm]
{\bf Acknowledgements}

This work was supported in part by DARPA under agreement nos.
MDA972-95-3-0008 and MDA972-96-3-014.


\newpage

Invited talk at the Second International Conference on Nitride
Semiconductors, Tokushima, Japan, October 27-31, 1997;
 accepted for publication in J. Cryst. Growth.

\end {document}